\documentclass[12pt]{article}
\usepackage{amsmath}
\usepackage{amssymb}
\usepackage{epsfig}
\usepackage{euscript}
\usepackage{fancybox}
\usepackage{color}
\def\mathswitchr#1{\relax\ifmmode{\mathrm{#1}}\else$\mathrm{#1}$\fi}

\newcommand {\pslash}{\hbox{$\not\hbox{\kern-2.3pt $p$}$}}

\usepackage{cite}

\usepackage{epic}
\def\alf1{ {\alpha\over\pi} }
\begin{document}
%\input{feynman} 
%=======================================================================
\begin{titlepage}
\begin{flushright}
%{\bf MPI-PhT-2002-08}\\
{\bf BU-HEPP-06-05 }\\
{\bf Apr., 2006}\\
\end{flushright}
%\vspace{0.05cm}
 
\begin{center}
{\Large Planck Scale Remnants in Resummed Quantum Gravity$^{\dagger}$
}
\end{center}

\vspace{2mm}
\begin{center}
%%  {\bf   S. Jadach$^{a,b}$ and B.F.L. Ward$^{c,d}$}
{\bf   B.F.L. Ward}\\
\vspace{2mm}
%{\em $^a$CERN, Theory Division, CH-1211 Geneva 23, Switzerland,}\\
%{\em $^b$Institute of Nuclear Physics,
%        ul. Kawiory 26a, Krak\'ow, Poland,}
%{\em $^c$Werner-Heisenberg-Institut, Max-Planck-Institut fuer Physik,
%Muenchen, Germany,}\\
%{\em $^a$Werner-Heisenberg-Institut, Max-Planck-Institut fuer Physik,
%Muenchen, Germany,}\\
{\em Department of Physics,\\
 Baylor University, Waco, Texas, 76798-7316, USA}\\
%{\em $^c$SLAC, Stanford University, Stanford, California 94309, USA,}\\
%{\em $^b$Department of Physics and Astronomy,\\
%  The University of Tennessee, Knoxville, Tennessee 37996-1200, USA}\\
%{\em $^c$SLAC, Stanford University, Stanford, California 94309, USA,}\\
\end{center}

\vspace{5mm}
\begin{center}
{\bf   Abstract}
\end{center}
We show that, in a new approach to quantum gravity in which its UV
behavior is tamed by resummation of large IR effects,
the final state of the Hawking radiation for an originally
very massive black hole
is a Planck scale remnant which
is completely accessible to our universe.
This remnant would be expected to decay into n-body final
states, leading to Planck scale cosmic rays.\\
\vskip 20mm
\centerline{Invited talk presented at the 2006 Cracow Epiphany Conference}
\vspace{10mm}
\renewcommand{\baselinestretch}{0.1}
\footnoterule
\noindent
{\footnotesize
\begin{itemize}
\item[${\dagger}$]
Work partly supported by US DOE grant DE-FG02-05ER41399 and 
% the Polish Government
%grants KBN 2P30225206 and 2P03B17210, the Maria Sk\l{}odowska-Curie
%Joint Fund II PAA/DOE-97-316, and
by NATO Grant PST.CLG.980342.
%, and by
%Polish Government grant 5P03B09320.
\end{itemize}
}
%\vspace{0.5cm}
%\begin{flushleft}
%{\bf UTHEP-00-0101}\\
%{\bf Jan, 2000}\\
%\end{flushleft}

\end{titlepage}

%=======================================================================
\def\Kmax{K_{\rm max}}\def\ieps{{i\epsilon}}\def\rQCD{{\rm QCD}}
\renewcommand{\theequation}{\arabic{equation}}
\font\fortssbx=cmssbx10 scaled \magstep2
\renewcommand\thepage{}
%\vfill\eject
\parskip.1truein\parindent=20pt\pagenumbering{arabic}\par
\section{\bf Introduction}\label{intro}\par
Recently~\cite{bw1,bw2,bw3,bw4}, we have introduced a new approach to
the problem of quantum general relativity, resummed quantum gravity, in which
we arrive at a UV finite representation of that theory. We start
from the original approach of Feynman in Refs.~\cite{ff1,ff2},
in which he showed how one represents Einstein's theory as
a sum of Feynman graphs by expanding in powers of $\kappa=\sqrt{8\pi G_N}$
where $G_N$ is Newton's constant. We improve on Feynman's results
by resumming the large infrared (IR) parts of these graphs using the extension
to non-Abelian IR algebras~\cite{qcdexp} of the methods of Ref.~\cite{yfs},
which we have used successfully in Refs.~\cite{sjward} in the Abelian case 
in which they were derived. 
We have found that this resummation tames the bad UV behavior found by
Feynman in Refs.~\cite{ff1,ff2}, so that the resulting resummed theory
is UV finite. In what follows, we apply this new UV finite approach to
quantum general relativity(QGR) to the problem of the final state of Hawking
radiation~\cite{hawk1} for an originally very massive black hole solution of Einstein's theory~\cite{mtw,sw1}.\par
We emphasize that our theory is one of several efforts to solve
the outstanding problem of quantum general relativity. The most successful
effort theoretically to date is of course the popular 
superstring theory~\cite{gsw,jp}. There is also the loop quantum gravity
approach~\cite{lpqg1} and the asymptotic safety approach~\cite{wein1} 
as realized phenomenologically in Refs.~\cite{lausch,reuter2,litim,perc} 
using the exact renormalization group apparatus. We also mention the low energy
effective Lagrangian approach~\cite{wein1} in which a low energy expansion
of the theory is made~\cite{dono1,lowenrest} in complete 
analogy with the successful
chiral perturbation theory low energy expansion in QCD~\cite{qcd1}.
An alternative approach to the IR regime for gravitational effects is
given as well in Refs.~\cite{sola}.
As we have explained in Refs.~\cite{bw1,bw2,bw3,bw4}, our approach in
no way contradicts any of these other important efforts.\par
Specifically, the since the new theory of QGR
is still in its early stages
of development and application, we start in the next section with a brief
review of the Feynman approach upon which the new theory is based.
This is followed in Section~\ref{rqgr} with a presentation of the elements
of the new theory. In Section~\ref{mbhs-hawk}, we then apply
the new theory to the problem of the final state of Hawking radiation
for an originally massive black hole solution of Einstein's theory
and show that it leads to Planck scale remnants.
Section~\ref{conc} contains some concluding remarks.\par
\section{Review of Feynman's Approach to QGR}
We specialize the Standard Model(SM) to its Higgs sector with just the
Higgs and its gravitational interactions, as this will allow us to
represent Feynman's approach to QGR. Any generalization to the spinning
particles in the SM can then be carried-out as needed by known
methods~\cite{bw1,bw2,bw3,bw4}. We take the mass of the Higgs
to be 120 GeV for definiteness, in view of the LEP data~\cite{lewwg}.
The relevant Lagrangian, already considered by Feynman in Refs.~\cite{ff1,ff2}, is then
{\small
\begin{equation}
\begin{split}
{\cal L}(x) &= -\frac{\sqrt{-g}}{2\kappa^2} R
            + \frac{\sqrt{-g}}{2}\left(g^{\mu\nu}\partial_\mu\varphi\partial_\nu\varphi - m_o^2\varphi^2\right)\\
            &= \frac{1}{2}{\big\{} h^{\mu\nu,\lambda}\bar h_{\mu\nu,\lambda} - 2\eta^{\mu\mu'}\eta^{\lambda\lambda'}
\bar{h}_{\mu_\lambda,\lambda'}\eta^{\sigma\sigma'}\\
&\bar{h}_{\mu'\sigma,\sigma'}{\big\}}
          + \frac{1}{2}{\big\{}\varphi_{,\mu}\varphi^{,\mu}-m_o^2\varphi^2 {\big\}} \\
&-\kappa {h}^{\mu\nu}{\big[}\overline{\varphi_{,\mu}\varphi_{,\nu}}+\frac{1}{2}m_o^2\varphi^2\eta_{\mu\nu}{\big{]}}\\
            & \quad - \kappa^2 [ \frac{1}{2}h_{\lambda\rho}\bar{h}^{\rho\lambda}{\big{(}} \varphi_{,\mu}\varphi^{,\mu} - m_o^2\varphi^2 {\big{)}} \\
&- 2\eta_{\rho\rho'}h^{\mu\rho}\bar{h}^{\rho'\nu}\varphi_{,\mu}\varphi_{,\nu}] + \cdots \\
\end{split}  
\label{eq1}
\end{equation}}\noindent
where $\varphi_{,\mu}\equiv \partial_\mu\varphi$ and 
we have the metric
$g_{\mu\nu}(x)=\eta_{\mu\nu}+2\kappa h_{\mu\nu}(x)$ with
$\eta_{\mu\nu}={\text diag}\{1,-1,-1,-1\}$ and 
$R$ is the curvature scalar. We define
%%%%{\Color{Black} $\kappa=\sqrt{8\pi {\Color{PineGreen}G_N}}$}
$\bar y_{\mu\nu}\equiv \frac{1}{2}\left(y_{\mu\nu}+y_{\nu\mu}-\eta_{\mu\nu}{y_\rho}^\rho\right)$ for any tensor $y_{\mu\nu}$.
%The Feynman rules for (\ref{eq1}) were
The Feynman rules for this theory were 
already worked-out by Feynman~\cite{ff1,ff2}.
where we use his gauge, $\partial^\mu \bar h_{\nu\mu}=0$. 
On this view,
%\item {\Color{Brown}$\varphi$} is representative of matter at a high scale 
%{\Color{Brown}$\sim M_{GUT}=10^{16}$GeV}
quantum gravity is just another quantum field theory
where the metric now has quantum fluctuations as well.
\par 

\section{Resummed QGR}
\label{rqgr}
We apply our extension in Ref.~\cite{qcdexp} of the methods of Ref.~\cite{yfs} 
to quantum general relativity, so that for the 
two point function for $\varphi$ we get~\cite{bw1}
\begin{equation}
i\Delta'_F(k)|_{\text{resummed}} =  \frac{ie^{B''_g(k)}}{(k^2-m^2-\Sigma'_s+i\epsilon)}
\label{resum}
\end{equation}
for{\small ~~~($\Delta =k^2 - m^2$)
\begin{equation}
\begin{split} 
B''_g(k)&= -2i\kappa^2k^4\frac{\int d^4\ell}{16\pi^4}\frac{1}{\ell^2-\lambda^2+i\epsilon}\\
&\qquad\frac{1}{(\ell^2+2\ell k+\Delta +i\epsilon)^2}\\
&=\frac{\kappa^2|k^2|}{8\pi^2}\ln\left(\frac{m^2}{m^2+|k^2|}\right),       
\end{split}
\label{yfs1} 
\end{equation}}
where the latter form holds for the UV regime, so that (\ref{resum}) 
falls faster than any power of $|k^2|$. An analogous result~\cite{bw1} holds
for m=0. We also note that, as $\Sigma'_s$ starts in ${\cal O}(\kappa^2)$,
we may drop it in calculating one-loop effects. It follows that
when the respective analogs of (\ref{resum}) are used, one-loop 
corrections are finite. In fact, it can be shown that the use of
our resummed propagators renders all quantum 
gravity loops UV finite~\cite{bw1}. We have called this representation
of the quantum theory of general relativity resummed quantum gravity (RQG).
\par
\section{Black Holes and Hawking Radiation to Planck Scale Remnants}
\label{mbhs-hawk}
The finiteness of the loop corrections in resummed quantum gravity (RQG)
allows us to address any number of outstanding problems in quantum general
relativity. Among those is the final state of the Hawking radiation
for an originally very massive black hole solution of Einstein's
theory. We now present our analysis of this problem.\par

\begin{figure}
\begin{center}
\epsfig{file=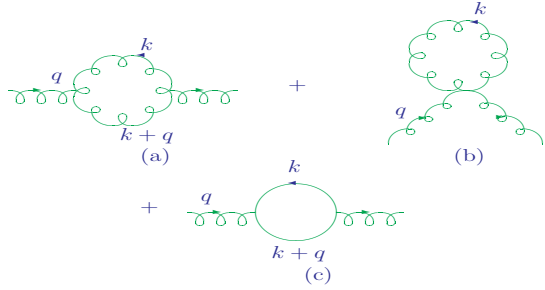,width=77mm,height=38mm}
\end{center}
\caption{\baselineskip=7mm  The graviton((a),(b)) and its ghost((c)) one-loop contributions to the graviton propagator. $q$ is the 4-momentum of the graviton.}
\label{fig1}
\end{figure}

\begin{figure}
\begin{center}
\epsfig{file=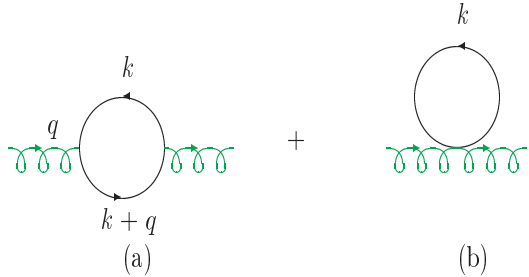,width=77mm,height=38mm}
\end{center}
\caption{\baselineskip=7mm  The scalar one-loop contribution to the
graviton propagator. $q$ is the 4-momentum of the graviton.}
\label{fig2}
\end{figure}

Specifically, we start by calculating the effects of the one-loop
corrections to the graviton propagator in Figs.~\ref{fig1} and \ref{fig2}. 
When the graphs Figs. 1 and 2 are computed 
in our resummed quantum gravity
theory as presented in Refs.~\cite{bw1,bw2,bw3}, the coefficient $c_{2,eff}$
in eq.(12) of Ref.~\cite{bw3}, which describes the attendant effect
on the denominator of the graviton propagator, becomes here, summing over the SM particles
in the presence of the
recently measured small 
cosmological constant~\cite{cosm1}, which implies the
gravitational infrared cut-off of $m_g\cong 3.1\times 10^{-33}$eV,   
\begin{equation}
c_{2,eff}=\sum_j n_j I(\lambda_c(j))
\label{ceff}
\end{equation} 
where we define~\cite{bw3} $n_j$ as the effective number of degrees of freedom
for particle $j$ and the integral $I$ is given by
\begin{equation}
I(\lambda_c)\cong \int^{\infty}_{0}dx x^3(1+x)^{-4-\lambda_c x}
\end{equation}
with the further definition $\lambda_c(j)=\frac{2m_j^2}{\pi M_{Pl}^2}$
where the value of $m_j$ is the rest mass of particle $j$
when that is nonzero. When the rest mass of particle $j$ is zero,
the value of $m_j$ turns-out to be~\cite{elswh} 
$\sqrt{2}$ times the gravitational infrared cut-off 
mass~\cite{cosm1}. We further note that, from the
exact one-loop analysis of Ref.\cite{thvelt1}, it also follows
that the value of $n_j$ for the graviton and its attendant ghost is $42$.
For $\lambda_c\rightarrow 0$, we have found the approximate representation
\begin{equation}
I(\lambda_c)\cong \ln\frac{1}{\lambda_c}-\ln\ln\frac{1}{\lambda_c}-\frac{\ln\ln\frac{1}{\lambda_c}}{\ln\frac{1}{\lambda_c}-\ln\ln\frac{1}{\lambda_c}}-\frac{11}{6}.
\end{equation} 
In this way, we obtain
from the standard Fourier transform of the respective graviton propagator
the improved Newton potential
\begin{equation}
\Phi_{N}(r)= -\frac{G_N M}{r}(1-e^{-ar}),
\label{newtnrn}
\end{equation}
where now, with 
\begin{equation}
c_{2,eff} \cong 2.56\times 10^{4}
\end{equation}
and, from eq.(8) in Ref.~\cite{bw3},
\begin{equation}
a \cong (\frac{360\pi M_{Pl}^2}{c_{2,eff}})^{\frac{1}{2}}
\end{equation}
we have that
\begin{equation}
a \cong  0.210 M_{Pl}.
\end{equation}
\par
What this means is that, when we make contact with the analysis
of Ref.~\cite{reuter2} for the final state of the Hawking
radiation for an originally very massive black hole, we can have a smooth
transition from the effective running Newton's constant $G(r)$ found
in Ref.~\cite{reuter2},
\begin{equation}
G(r)=\frac{G_Nr^3}{r^3+\tilde{\omega}G_N\left[r+\gamma G_N M\right]}
\label{rnG}
\end{equation}
for a central body of mass $M$ where $\gamma$ is a phenomenological
parameter~\cite{reuter2} satisfying $0\le\gamma\le\frac{9}{2}$ and
$\tilde{\omega}=\frac{118}{15\pi}$, to our result in (\ref{newtnrn}) via the
matching at the outermost solution, $r_>$, of the equation
\begin{equation}
G(r)=G_N(1-e^{-ar}).
\label{match}
\end{equation}
For $r<r_>$, in the metric class 
\begin{equation}
ds^2 = f(r)dt^2-f(r)^{-1}dr^2 - r^2d\Omega^2,
\end{equation}
the lapse function is then taken to be
\begin{equation}
%\begin{split}
f(r) =1-\frac{2G_{eff}(r)M}{r}
%    &= \frac{B(x)}{B(x)+2x^2}|_{x=\frac{r}{G_NM}},
%\end{split}
\label{reuter1}
\end{equation}
with $G_{eff}(r)=G_N(1-e^{-ar})$ whereas for $r>r_>$ we set 
$G_{eff}(r)=G(r)$. It the follows~\cite{bw4} that for the self-consistent 
value $\gamma=0$ and $0.2=\Omega\equiv\frac{\tilde\omega}{G_NM^2}=\frac{\tilde\omega M_{Pl}^2}{M^2}$ for definiteness we find that the 
inner horizon found in Ref.~\cite{reuter2}
moves to negative values of $r$ and 
that the outer horizon moves to $r=0$, so that
the entire mass of the originally very massive black hole radiates away
until a Planck scale remnant of mass $M_{cr}'=2.38~M_{Pl}$ is left\footnote{Ref.~\cite{maart} argues as well that the loop
quantum gravity approach implies that black holes below a critical
mass do not form, in agreement with Ref.~\cite{reuter2}. Here, we show that the attendant remnants are actually accessible to our universe.},
which then is completely accessible to our universe. It would be expected
to decay into n-body final states, $n=2,3,\cdots$, leading in general
to Planck scale cosmic rays. The data in Ref.~\cite{cosmicray,westerhoff} 
are not inconsistent with this conclusion, which also agrees with
recent results by Hawking~\cite{hawk2}.\par

\section{Conclusions}
\label{conc}
We conclude that resummed quantum gravity allows us to address many 
questions that have hitherto been either intractable or very cumbersome 
in rigorous quantum field theory. In our discussion, we have argued that
the final state of the Hawking radiation from an originally
very massive black hole is a Planck scale remnant that
is entirely accessible to our universe and that would be expected
to decay into Planck scale cosmic rays. We encourage experimentalists
to search for such.\par

\section*{Acknowledgments}

We thank Prof. S. Jadach for useful discussions.

\end{document}